\documentclass[%
 reprint,
superscriptaddress,
nofootinbib,
 amsmath,amssymb,
 aps,
prb,
]{revtex4-2}

\usepackage{graphicx}
\usepackage{dcolumn}
\usepackage{bm}
\usepackage{soul}
\usepackage{longtable}
\usepackage{booktabs}
\usepackage{multirow}
\usepackage{subfigure}
\usepackage{amsfonts}
\usepackage{amssymb}
\usepackage{amsmath}
\usepackage{color}
\usepackage{lipsum}
\usepackage{braket}
\usepackage{bbm}

\definecolor{LinkColor}{RGB}{192,77,77}
\usepackage{hyperref}
\hypersetup{
	pdfauthor={good guys},
	pdftitle={good title},
	colorlinks=true,
	citecolor=LinkColor,
	linkcolor=LinkColor,
	urlcolor=LinkColor
}

\begin{document}

\preprint{APS/123-QED}
\title{Fate of non-Hermitian free fermions with Wannier-Stark ladder}

\author{Han-Ze Li}
\affiliation{Institute for Quantum Science and Technology (IQST)\\ and Department of Physics, Shanghai University, Shanghai 200444, China}

\author{Minhui Wan}
\affiliation{Institute for Quantum Science and Technology (IQST)\\ and Department of Physics, Shanghai University, Shanghai 200444, China}
\affiliation{School of Physics and Optoelectronics, Xiangtan University, Xiangtan 411105, China}

\author{Jian-Xin Zhong}
\email{jxzhong@shu.edu.cn}
\affiliation{Institute for Quantum Science and Technology (IQST)\\ and Department of Physics, Shanghai University, Shanghai 200444, China}
\affiliation{School of Physics and Optoelectronics, Xiangtan University, Xiangtan 411105, China}

\date{\today}

\begin{abstract}
The Wannier-Stark localization dynamically alters the entanglement behavior of non-Hermitian free fermions. Utilizing the single-particle correlation matrix technique, we analyze the effective Hamiltonian of these fermions with a Wannier-Stark ladder. Under open boundary conditions, we observe the steady state half-chain entanglement entropy and identify two distinct area law regions and an algebraic scaling region. Finite-size scaling analysis reveals critical scaling behavior of the half-chain entanglement entropy. Notably, the system demonstrates unique entanglement characteristics under periodic boundary conditions, which diverge from the $(1+1)d$ conformal field theory predictions for Anderson localization. Our findings highlight novel entanglement phases emerging from the interplay between the non-Hermitian skin effect and disorder-free localization.
\end{abstract}

\maketitle
\section{Introduction}
The interplay between unitary evolution and measurement in quantum many-body systems forms a complex dynamical equilibrium, which could lead to a rich pattern of entanglement and correlation dynamics. The unitarity of evolution ensures probability conservation and the linearity of the Schrödinger equation, which constrain the spread of information and correlations in an isolated quantum many-body system. However, the continually unitary evolution can be punctuated by projective measurements, resulting in disentanglement and decoherence of the system. In this context, the nonunitary evolution introduced by projective measurement induces entanglement scaling transitions between volume-law phase ($S\propto L^{d}$) and area-law quantum Zeno phase ($S\propto L^{d-1}$)~\cite{area_law}. Measurement-induced entanglement transitions (MIET) are evident in the frameworks of quantum Hamiltonians and monitored quantum circuits~\cite{1,2,3,4,5,6,7,8,9,10,11,12,13,14,15,16,17,18,19,20,21,22, 10.21468/SciPostPhys.15.4.170, chen2023robust}. This has unveiled a quantum information perspectives on condensed matter systems.

The effective non-Hermitian Hamiltonian~\cite{45,46}, which evolves under continuous measurement without quantum jumps~\cite{47}, could also serve as a viable setup for studying the MIET in nonunitary evolution scenarios~\cite{47,48}. The most interesting case is the interplay between the non-Hermitian skin effect (NHSE)~\cite{23,24,25,26,27,28,29,30,31,32,33,34,35,36,37,38,39,40,41,szl,43,44}, characterized by a large number of eigenmodes being anomalously localized at one side of the open boundary due to nonreciprocal dissipation, and a free fermion system~\cite{49,50}. There, weak measurements and post-selections encapsulated by the non-Hermicity~\cite{47}, cause non-Hermitian free fermions to undergo a volume-to-area law entanglement scaling transition. Subsequently,~\cite{51} exploring how the interaction between the NHSE in non-Hermitian free fermions and Anderson localization~\cite{Anderson} leads to divergent scaling laws in log-to-volume-to-area phases under non-Hermitian disordered evolution. The similar phenomenon was also observed in the non-Hermitian Aubry-André-Harper chain~\cite{52}, albeit with entanglement transitions that follow area-to-volume-to-area scaling laws. Meanwhile, transitions between typical volume-to-area scaling laws and area-to-log-to-volume scaling laws in two distinct non-Hermitian quasiperiodic systems were confirmed ~\cite{53}. In addition, these similar physical issues, entanglement dynamical competition between localization and NHSE, are discussed in the context of disordered monitored free fermions \cite{54}. Clarifying the intricate entanglement dynamics between the NHSE and localization significantly enhances our understanding of the MIET during nonunitary evolution.

\begin{figure}[bt]
	\includegraphics[width=8.2cm]{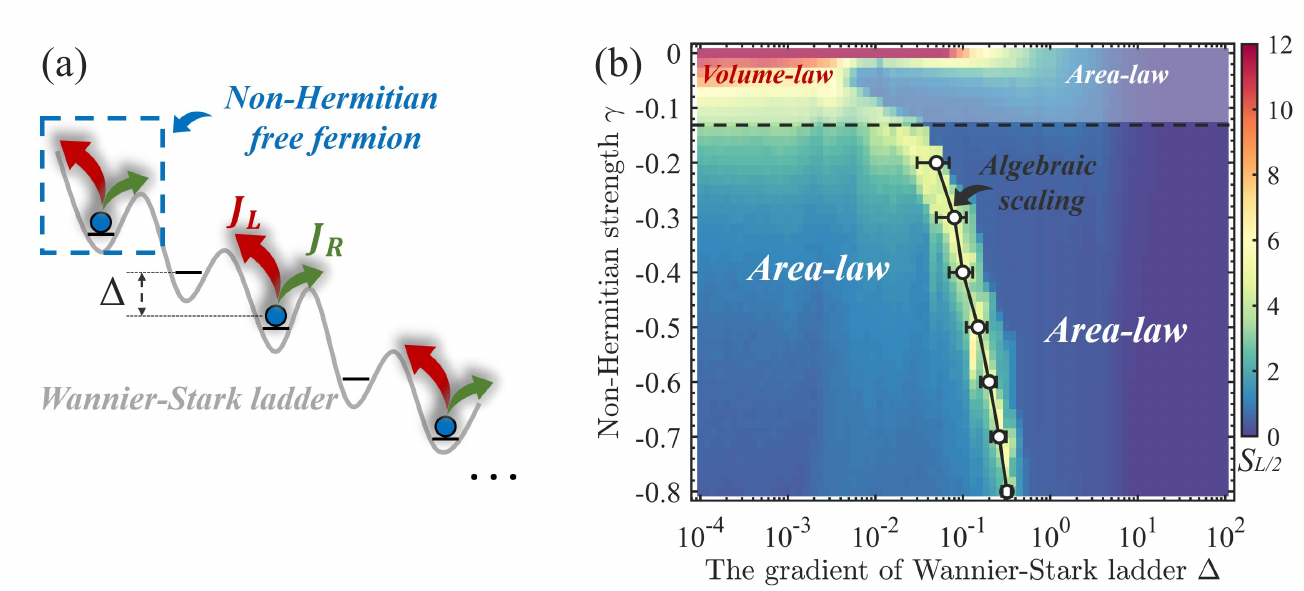}
	\caption{\label{fig1} (a) Schematic representation of non-Hermitian free fermions WSL. (b) Entanglement phase diagram as a function of non-Hermitian strength $\gamma$ and WSL gradient $\Delta$, with system size $L=320$. The color density plot represents the estimated half-chain entanglement entropy $S_{L/2}$. The transition boundary is inferred from data collapse of the half-chain EE, shown by the black scatter plot.}
\end{figure}

However, the entanglement properties in non-Hermitian free fermions within disorder-free localization systems, particularly non-Hermitian free fermions with Wannier-Stark ladder (WSL)~\cite{WSL}, remain largely unexplored. The WSL shows that a constant electric field causes electrons in a lattice to have discrete, localized energy levels, forming a ladder-like structure that significantly affects their dynamics and transport properties \cite{55,56,57,58,59,A1,A2,A3,A4,A5,A6,WSL}. The pioneering work by M. Fisher~\cite{Fisher} and the systematic studies by Andrey Kolovsky~\cite{A1,A2,A3,A4,A5,A6,WSL} on the Wannier-Stark system, have laid a crucial foundation for understanding these effects, including transport and localization phenomena. Notably, due to its disorder-free nature, the WSL holds nontrivial experimental and numerical advantages. Naturally, what are the entanglement dynamics of non-Hermitian free fermions with WSL? What is the interplay between many-body NHSE and WSL?

To address the aforementioned issues, in this work, we investigate the entanglement dynamics and critical scaling behaviors in the non-Hermitian free fermions with WSL. Using the single-particle correlation matrix technique, we observe that under open boundary condition (OBC), the system undergoes a rich entanglement transition as the gradient of WSL increases at higher non-Hermitian strength. This transition ranges from the area law phase (induced by the NHSE) to algebraic scaling phase, and finally re-entry to the area law phase (induced by Wannier-Stark localization). This indicates that the inclusion of WSL enriches the entanglement dynamics of non-Hermitian free fermions. Additionally, under periodic boundary condition (PBC), the entanglement behavior only exhibits a transition from volume law phase to area law phase.

The rest of the paper is organized as follows. In Sec.~\ref{model and methods}, we first introduce the effective Hamiltonian of non-Hermitian free fermions with WSL and the numerical methods employed. In Sec.~\ref{Results and discussions}, we present the numerical results to identify the physical quantities necessary for detecting entanglement phase diagram and transitions. We also give a brief discussion from the perspective of single-particle localization in Sec.~\ref{Discussions}. Conclusion and outlooks are given in Sec.~\ref{conclusion_and_outlooks}. Additional numerical calculation data are included in the Appendix.

\section{Model and methods}\label{model and methods}
\subsection{Effective Hamiltonian of Non-Hermitian free fermions model with Wannier-Stark ladder}
In this work, we consider an effective Hamiltonian [see Appendix~\ref{nh-effective H} for the discussion of the effective non-Hermitian Hamiltonians] of one-dimensional spinless free fermions with Hatano-Nelson nearest neighbor (NN) hopping, subject to a WSL [see FIG.~\ref{fig1}(a)]:
\begin{align}\label{eq1}
    H_{\rm eff} = \sum ^{L-1} _ { j = 1 }  ( J _ { L } c _ { j } ^ { \dagger } c _ { j + 1 } + J _ { R } c _ { j + 1 } ^ { \dagger} c _ { j } ) + \sum ^{L} _ { j = 1 } F _ { j } n_j ,
\end{align}
where the Hatano-Nelson NN hopping amplitudes are denoted by $J _ { L } \equiv - (1 - \gamma)$ and $J _ { R } \equiv - (1 + \gamma)$, where $\gamma$ represents the asymmetric non-Hermitian strength. The on-site potential at site index $j$ is given by
\begin{align}
    F_j \equiv \Delta \cdot j,
\end{align}
indicating a WSL, $\Delta$ is the gradient of WSL and $L$ is the finite system size, ensuring $L$ is even. With no WSL ($\Delta=0$) and OBCs, Eq.~\ref{eq1} demonstrates NHSE when $J_L\neq J_R\neq 0$, resulting in most many-body eigenstates tending to localize at a single boundary.

In the following, we set the system initially begin in a separable $\mathbb{Z}_2$ state given by $\ket{\psi_0} = \prod_{j=1}^{L/2} c^\dagger_{2j} \ket{\rm vac}$, where $\ket{\rm vac}$ represents the fermionic vacuum state. The final state $\ket{\psi(t)}$, evolved and normalized from $\ket{\psi_0}$ under the effctive Hamiltonian $H_{\rm eff}$, is given by
\begin{align}
\ket{\psi(t)} = \frac{e^{-iH_{\rm eff
}t}\ket{\psi_0}}{\sqrt{\bra{\psi_0}e^{iH_{\rm eff
}t}e^{-iH_{\rm eff
}t}\ket{\psi_0}}}.
\end{align}

It is worth noting that, still the effective Hamiltonian $H_{\rm eff}$ is non-Hermitian, we can conserve the evolving number of particles through post-selected a trajectory for no quantum jump [see Appendix~\ref{nh-effective H} for more details]. Additionally, thanks to the quadraticity of the Hamiltonian in Eq.~\ref{eq1} allows us to efficiently calculate its dynamics from the single-particle correlation matrix technique [see Appendix~\ref{appendix_single-particle and EE} for details].

\subsection{Entanglement entropy and mutual information}\label{entangle and MI}
For one thing, the entanglement entropy (EE) of a pure many-body state in a system, which quantifies the level of quantum entanglement between a subsystem and its environment, is determined by the subsystem's reduced density matrix, $\rho_{re}$. To be concrete, let us consider the von Neumann EE, which is defined as 
\begin{align}
    S_A = -{\rm Tr}[\rho_A \log \rho_A],
\end{align}
where $\rho_A = {\rm Tr}_{\bar{A}}(\rho)$ represents the reduced density matrix of a subsystem, obtained by tracing over the environmental part $\bar{A}$. In non-Hermitian systems, due to the fact that the left and right eigenstates are not conjugates of each other, it is common to distinguish between two types of density matrices~\cite{60, 61}. The EE calculated using these two different density matrices will show quantitative differences but no qualitative distinctions. In this work, we focus on the EE calculated using the biorthogonal density matrix~\cite{density_matrix}, which provides a self-consistent theoretical framework for non-Hermitian quantum mechanics \cite{62}. The quadratic form of the effective Hamiltonian $H_{\rm eff}$ in Eq.~\ref{eq1}, ensures that the evolved state $\ket{\psi(t)}$ retains its Slater determinant structure, thereby simplifying the computation of its correlation matrix $C^A_{ij}(t) = \bra{\psi(t)}c^\dagger_i c_j\ket{\psi(t)}$. After evolving over a long time to reach a steady state, we have derived the final expression for the EE in the non-Hermitian free fermions scenario:
\begin{align}
    S_A = -\sum_k \lambda_k\log \lambda_k + (1-\lambda_k)\log(1-\lambda_k).
\end{align}
where $\lambda_k$ is the eigenvalue of the steady state correlation matrix $C^A_{ij}$. We define the subsystem size $A$ as the series $\{1, 2, 3,\cdots \ell$, and $\ell\leq L\}$ and refer to its steady state EE as $S_\ell$ in the following. For more details on the numerical simulation of the steady state EE see Appendix~\ref{appendix_single-particle and EE}.

For another, we use mutual information as a probe to further confirm the entanglement scaling phases, which is defined as
\begin{align}
    I_{A:B}=S_A+S_B-S_{AB}.
\end{align}
Here, $S_A$ and $S_B$ represent the steady state EE of subsystems $A$ and $B$ respectively, while $S_{AB}$ denotes the steady state EE of the combined subsystems $A \cup B$. Mutual information not only exhibits scaling that is consistent with the EE but also functions effectively as an additional metric for identifying the presence of conformal symmetry.

\section{Results}\label{Results and discussions}
\subsection{Phase diagram}\label{phasediagram}
Before delving into phase transitions, it's essential to first explore the various quantum phases present within a phase diagram. To begin, let us consider the entanglement behavior of the late-time steady state $\ket{\psi(t)}$ in specific limiting scenarios: When $\gamma=0$, the effective Hamiltonian $H_{\rm eff}$ reverts to the Hermitian case. At $\Delta=0$, it exhibits volume-law phase. While, as $\Delta$ becomes non-zero, the system transitions to area-law entanglement scaling due to Wannier-Stark localization \cite{55,56,57,58}. However, in the absence of WSL, the Hatano-Nelson type hopping ($\gamma \neq 0$) pushes the system into a different phase. This phase is characterized by NHES-induced area-law phase, resulting from the localization of all many-body eigenstates at the boundary, driven by macroscopic particle flow during long-time evolution.

Utilizing the results for the steady state half-chain EE $S_{L/2}$, we draw a global phase diagram dependence of $S_{L/2}$ on the strength of non-Hermicity $\gamma$ and the gradient of WSL $\Delta$ [see FIG.~\ref{fig1}(b)]. Our numerical results (up to $L=320$) indicate that an increasing gradient of WSL, $\Delta$, leads to a more rich phases structure. Specifically, at small non-Hermitian strengths $\gamma$, the system undergoes a conventional phase transition from volume-law to area-law entanglement driven by the WSL, as illustrated in the light white region of FIG~\ref{fig1}(b). Venturing further into regions of stronger non-Hermiticity $\gamma$, in areas with the small gradient of WSL $\Delta$, the steady state half-chain EE exhibits a low-entanglement area-law growth. As $\Delta$ increases, the system reaches a transient phase where the growth of EE displays an unusual algebraic scaling. Then, at even higher $\Delta$, the system re-entry a phase characterized by area-law behavior.

\begin{figure}[bt]
	\includegraphics[width=8.7cm]{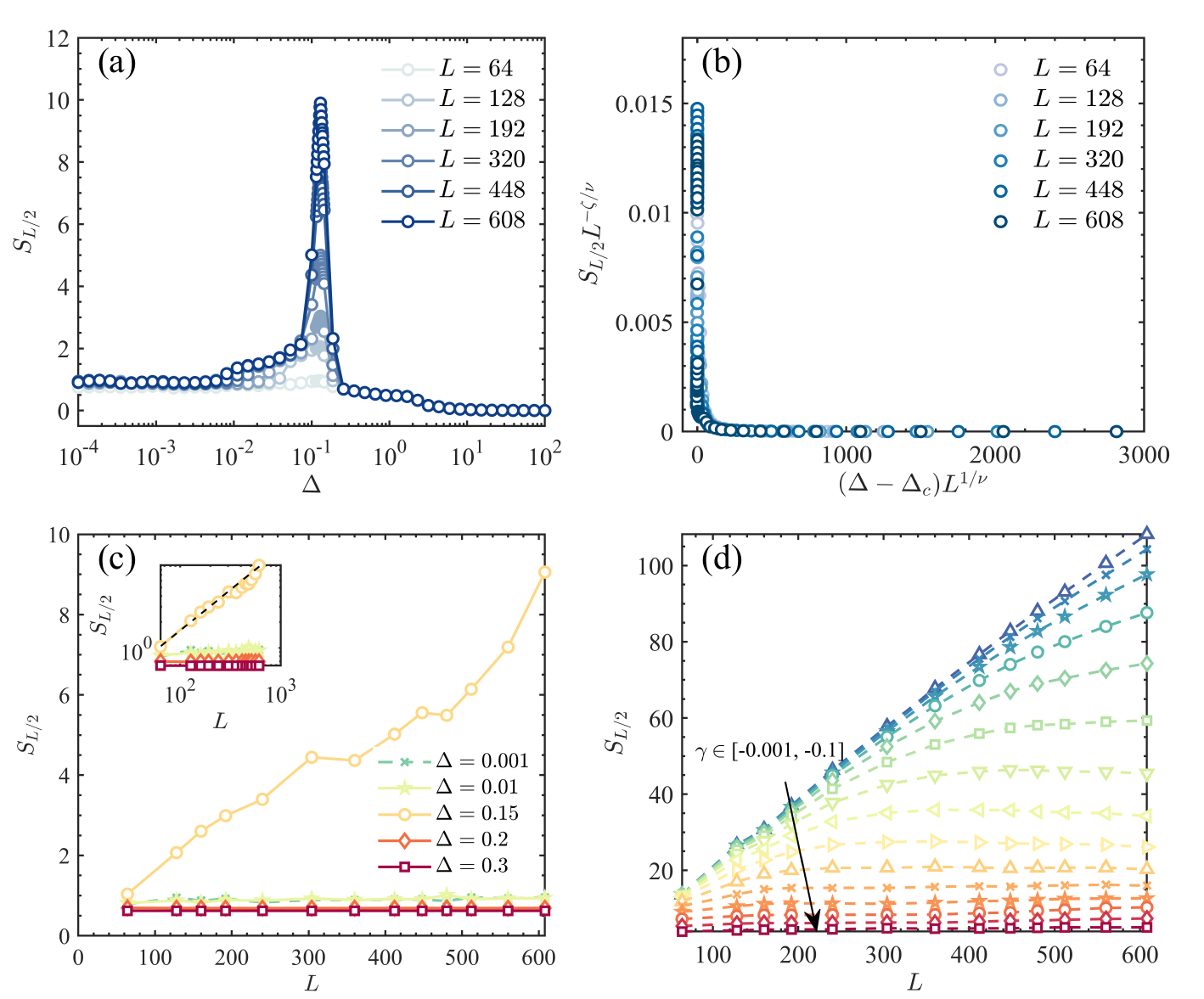}
	\caption{\label{fig2} (a) Plot of $S_{L/2}$ against the WSL gradient $\Delta$ for system sizes $L=64$, $128$, $192$, $320$, $448$, and $608$. (b) Data collapse of the steady-state $S_{L/2}$ using the scaling function derived in Eq.~\ref{eq7}, focusing on $\Delta \geq 0.1$ to achieve the collapse. (c) Variation of $S_{L/2}$ with different $\Delta$ values. The inset shows this data in a log-log scale, where the dashed line fits a power-law for $\Delta = 0.15$, described by $S_{L/2} \propto L^{0.922 \pm 0.215}$. (d) Dependence of $S_{L/2}$ on non-Hermitian strength $\gamma$ at a fixed $\Delta = 0.001$, analyzed under OBCs.}
\end{figure}

\subsection{Entanglement transitions and critical scaling properties}\label{EE transition}

To provide a deeper insight into the occurrence of entanglement phase transitions, we first examine the steady-state half-chain EE $S_{L/2}$. In FIG.~\ref{fig2}(a), $S_{L/2}$ is presented as a function of $\Delta$ for various system sizes $L$ at a fixed $\gamma = -0.5$. We observe that within a certain peak range of $\Delta$, the steady-state EE increases with system size $L$, while in other regions, the steady-state EE remains low and shows little dependence on the system size.

To identify the critical points of the transitions more precisely, we employ a finite-size scaling method~\cite{andreas_sorge_2015_35293} for $S_{L/2}$, as described by
\begin{align}
    S_{L/2}(\Delta) = L^{\zeta/\nu} f[L^{1/\nu}(\Delta - \Delta_c)],\label{eq7}
\end{align}
where $L \rightarrow \infty$ and $\Delta \rightarrow \Delta_c$. The scaling function $f(x)$ behaves as:
\begin{align}
    f(x) \propto \left\{ 
    \begin{array}{cc}
        \text{const.}, & x \rightarrow 0 \\
        |x|^{-\zeta}, & |x| \gg 1.
    \end{array}
    \right.
\end{align}

Using this scaling function from FIG.~\ref{fig2}(b) to collapse the $S_{L/2}$ data, we determine the critical point $\Delta_c \approx 0.15 \pm 0.04$, with critical scaling exponents $\zeta \approx 1.98 \pm 0.06$ and $\nu \approx 1.92 \pm 0.04$.

FIG.~\ref{fig2}(c) illustrates the half-chain EE $S_{L/2}$ as a function of system size $L$ for fixed $\gamma = -0.5$ and varying $\Delta$ values of 0.001, 0.01, 0.15, 0.2, and 0.3. It is evident that when $\Delta < \Delta_c$ or $\Delta > \Delta_c$, the half-chain EE $S_{L/2}$ exhibits area-law behavior, remaining approximately constant with respect to $L$. However, at the critical scaling point $\Delta = \Delta_c$, the half-chain EE $S_{L/2}$ shows algebraic growth with system size $L$.

The inset of FIG.~\ref{fig2}(c) fits the critical scaling exponent $S_{L/2} \sim L^\beta$, yielding $\beta \approx 0.922 \pm 0.215$, consistent with the finite-size scaling results with $\zeta / \nu \approx 1.03(1)$~\cite{scaling}. FIG.~\ref{fig2}(d) shows the transition from volume-law to area-law behavior as a function of system size $L$ under a small gradient of WSL ($\Delta = 0.001$), driven by increasing non-Hermitian strength $\gamma$ in the range $[-0.001, -0.1]$, due to the NHSE.

To further delineate the critical scaling lines $\{\Delta_c\}$ in the phase diagram FIG.~\ref{fig1}(b) as functions of $\gamma$ and $\Delta$, the finite-size scaling method previously described was used to calculate the critical points and exponents for $\gamma = -0.2, -0.3, -0.4, -0.5, -0.6, -0.7, -0.8$. It is worth noting that the average values of the critical exponents $\nu$ and $\zeta$ for different $\gamma$ values are $\bar{\nu} = 1.876$ and $\bar{\zeta} = 2.034$~\cite{average}. TABLE~\ref{table1} in Appendix~\ref{appendix_datacollapse} presents the numerical values for the critical scaling points and exponents for different $\gamma$ values, including error estimates based on data collapse. Additional data on finite-size scaling are provided in Appendices~\ref{appendix_engtanglement} and~\ref{appendix_datacollapse} for further details.

\subsection{Deviates from $(1+1)d$ conformal field theory}\label{CFT}

The behavior of the EE here inevitably brings to mind the entanglement properties of quantum critical systems described by conformal field theory (CFT). Logarithmic scaling of EE serves as a key method for unveiling the distinct quantum correlations and coherence that characterize quantum critical systems, thereby offering essential insights into quantum phase transitions and quantum information behavior. Typically, in one-dimensional conformally invariant quantum systems under OBCs, which encompass both gapless critical systems \cite{63,64,65} and continuously monitored systems \cite{2,66,67}, the EE usually follows:

\begin{align}
    S_l \sim \left(\frac{c}{6}\right)\log\left[\sin\left(\frac{\pi l}{L}\right)\right] + \text{const.} \label{S_l}
\end{align}

However, as shown below, non-Hermitian free fermions with the Wannier-Stark localization deviate from the predictions of $(1+1)d$ CFTs.

\begin{figure}[bt]
    \includegraphics[width=8.6cm]{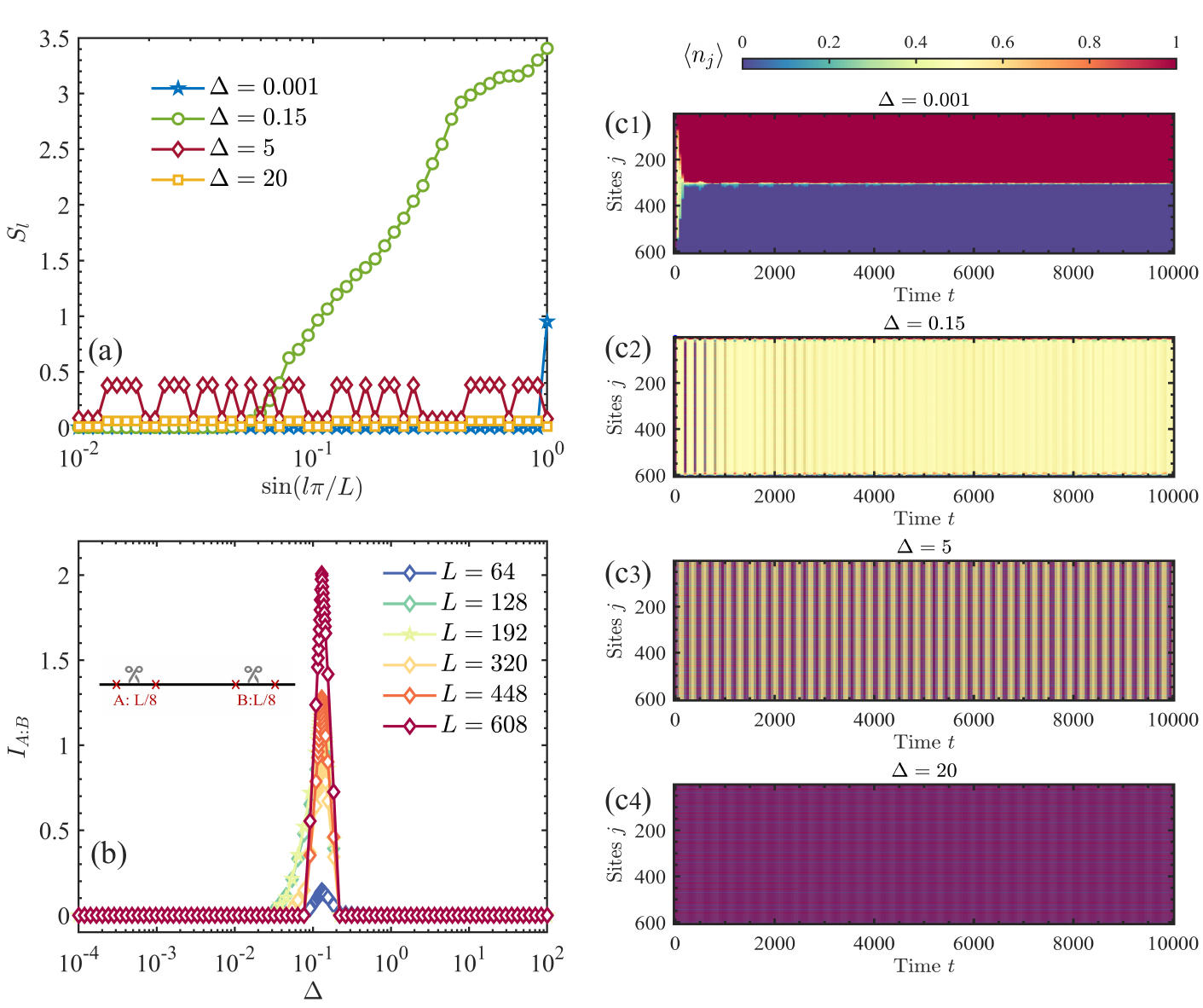}
    \caption{\label{fig:MI_Sl} (a) $S_l$ plotted for subsystems of varying lengths $l$ at WSL gradients $\Delta = 0.001$, $0.15$, $5$, and $20$. (b) Mutual information $I_{A:B}$ as a function of $\Delta$ for different system sizes $L=64$, $128$, $192$, $320$, $448$, and $608$, with $\gamma = -0.5$ under open boundary conditions (OBCs). (c1)-(c4) Time evolution of the density distribution for WSL gradients $\Delta = 0.001$, $0.15$, $5$, and $20$, with $\gamma = -0.5$, analyzed under OBCs.}
\end{figure}

First, we numerically computed the subsystem EE, $S_l$, at non-Hermitian strength $\gamma = -0.5$, as shown in FIG.~\ref{fig:MI_Sl}(a). When $\Delta = 0.001$, which is below the critical value $\Delta_c \approx 0.15$, $S_l$ is zero for most subsystem sizes $l$ but shows a nonzero value near $\sin(\pi l/L) = 1$ [see the light blue star-dotted line in FIG.~\ref{fig:MI_Sl}(a)]. This deviates from the logarithmic scaling described by Eq.~\ref{S_l}. This behavior is related to a single domain-wall in the density profile $\langle n_j \rangle$ caused by the NHSE. Specifically, when $l$ spans the domain-wall (i.e., $\langle n_j \rangle = 1$ or $0$), $S_l$ is nearly zero because the reduced density matrix $\rho_A$ approximates a pure state. 

At the critical scaling value $\Delta_c \approx 0.15$, the EE $S_l$ exhibits algebraic scaling with increasing subsystem size $l$, as indicated by the light green dotted line in FIG.~\ref{fig:MI_Sl}(a). When $\Delta = 5$ and $\Delta = 20$, which exceed the critical value $\Delta_c$, the subsystem EE $S_l$ shows area-law scaling due to Wannier-Stark localization.

In FIG.~\ref{fig:MI_Sl}(b), we present the mutual information $I_{A:B}$ for the system under OBC with $\gamma = -0.5$, where non-adjacent subsystems $A$ and $B$ are chosen, each of length $L/8$. The results reveal a pronounced peak in the mutual information near the critical scaling point $\Delta_c \approx 0.15$ as the system size $L$ increases, while displaying zero values in other regions. This indirectly describes the loss of independence between non-adjacent subsystems $A$ and $B$ near $\Delta_c$, and the preservation of independence away from $\Delta_c$. This contrasts with conformally invariant systems, where the mutual information typically maintains a non-zero constant value.

Furthermore, we calculated the time-evolution of density distributions for $\Delta < \Delta_c$, $\Delta = \Delta_c$, and $\Delta > \Delta_c$, as shown in FIG.~\ref{fig:MI_Sl}(c1)-(c4) with $\gamma = -0.5$. When $\Delta = 0.001 < \Delta_c \approx 0.15$, the NHSE causes particle accumulation on one side of the system, creating a single domain-wall, as shown in FIG.~\ref{fig:MI_Sl}(c1). At the critical scaling value $\Delta_c \approx 0.15$, the long-time propagation of particles causes the EE to grow, leading to thermalization of the density profile over time, as shown in FIG.~\ref{fig:MI_Sl}(c2). However, for $\Delta = 5$ and $\Delta = 20 > \Delta_c \approx 0.15$, Wannier-Stark localization restricts particle transport, resulting in low or even zero growth of the EE ($\Delta = 20$), thus preserving the initial state information to a significant extent.

\subsection{Non-Hermitian free-fermions with Wannier-Stark ladder under periodic boundaries}\label{PBC}

\begin{figure}[bt]
    \includegraphics[width=8.6cm]{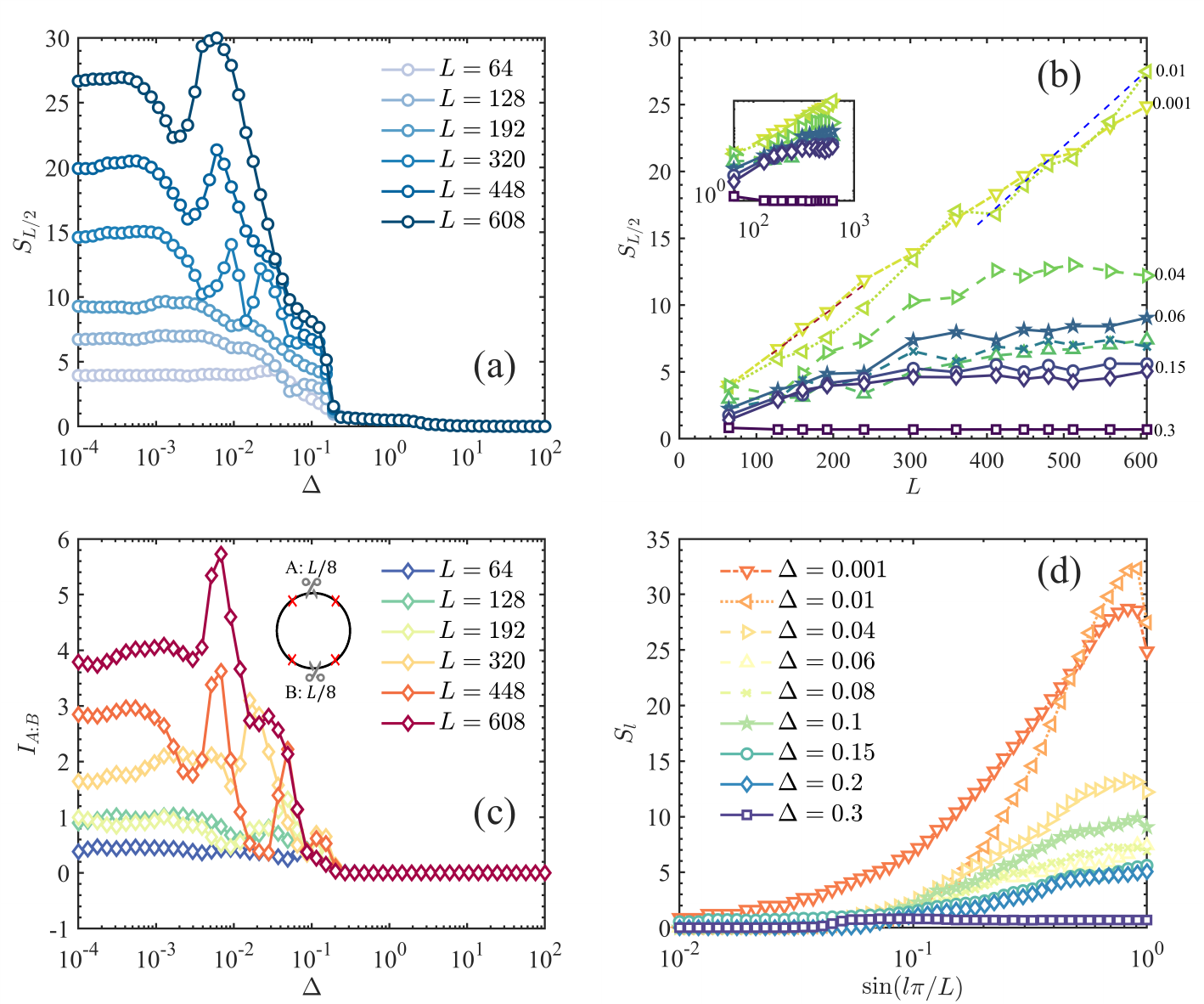}
    \caption{\label{fig:pbc} Entanglement properties under PBCs. (a) The steady-state half-chain entanglement entropy $S_{L/2}$ as a function of the WSL gradient $\Delta$ for system sizes $L=64$, $128$, $192$, $320$, $448$, and $608$. (b) The half-chain EE $S_{L/2}$ plotted against system size $L$ for $\Delta$ values ranging from $0.001$ to $0.3$. The inset shows $S_{L/2}$ in a log-log scale as a function of $\Delta$. (c) The mutual information $I_{A:B}$ between subsystems $A$ and $B$ (each of length $L/8$) plotted against $\Delta$ for different system sizes. (d) The entanglement entropy $S_l$ for subsystems of length $l$ for various $\Delta$.}
\end{figure}

We now discuss the entanglement behavior of Eq.~\ref{eq1} under PBCs with non-Hermitian strength $\gamma = -0.5$. In FIG.~\ref{fig:pbc}(a), we show the behavior of the half-chain entanglement entropy (EE) $S_{L/2}$ as a function of $\Delta$ for different system sizes $L$. For all sizes $L$, $S_{L/2}$ initially increases to a small peak and then sharply decreases as $\Delta$ increases. The peak shifts towards lower $\Delta$ values as the system size increases, suggesting that smaller $\Delta$ values are more effective in increasing EE in larger systems. High $\Delta$ values tend to localize the system, reducing EE.

In FIG.~\ref{fig:pbc}(b), we illustrate how $S_{L/2}$ depends on system size $L$ for $\Delta = 0.001 \sim 0.3$. As shown, both $\Delta = 0.001$ and $\Delta = 0.01$ exhibit volume law scaling behavior, while an increase in $\Delta$ leads to area law scaling. The fitted $\beta$ values for $\Delta = 0.001$ and $\Delta = 0.01$ are approximately $0.826 \pm 0.302$ and $0.903 \pm 0.251$, respectively.

FIG.~\ref{fig:pbc}(c) depicts the mutual information $I_{A:B}$ between the non-adjacent subsystems $A(L/8)$ and $B(L/8)$ under PBCs. The mutual information slightly peaks with $\Delta$ evolution, then decreases to zero as $\Delta$ further increases. Despite the presence of the small peak, the system exhibits a singular transition from volume-to-area law as $\Delta$ increases. Additionally, FIG.~\ref{fig:pbc}(d) illustrates that $S_l$, as a function of $\sin(\pi l/L)$, provides further evidence that with increasing $\Delta$, the system undergoes a sole entanglement phase transition from volume-law to area-law.

In summary, under PBCs, non-Hermitian free fermions with WSL behave differently from those with Anderson localization \cite{51,52,53}.

\section{Discussions}\label{Discussions}
\begin{figure}[bt]
	\includegraphics[width=6.6cm]{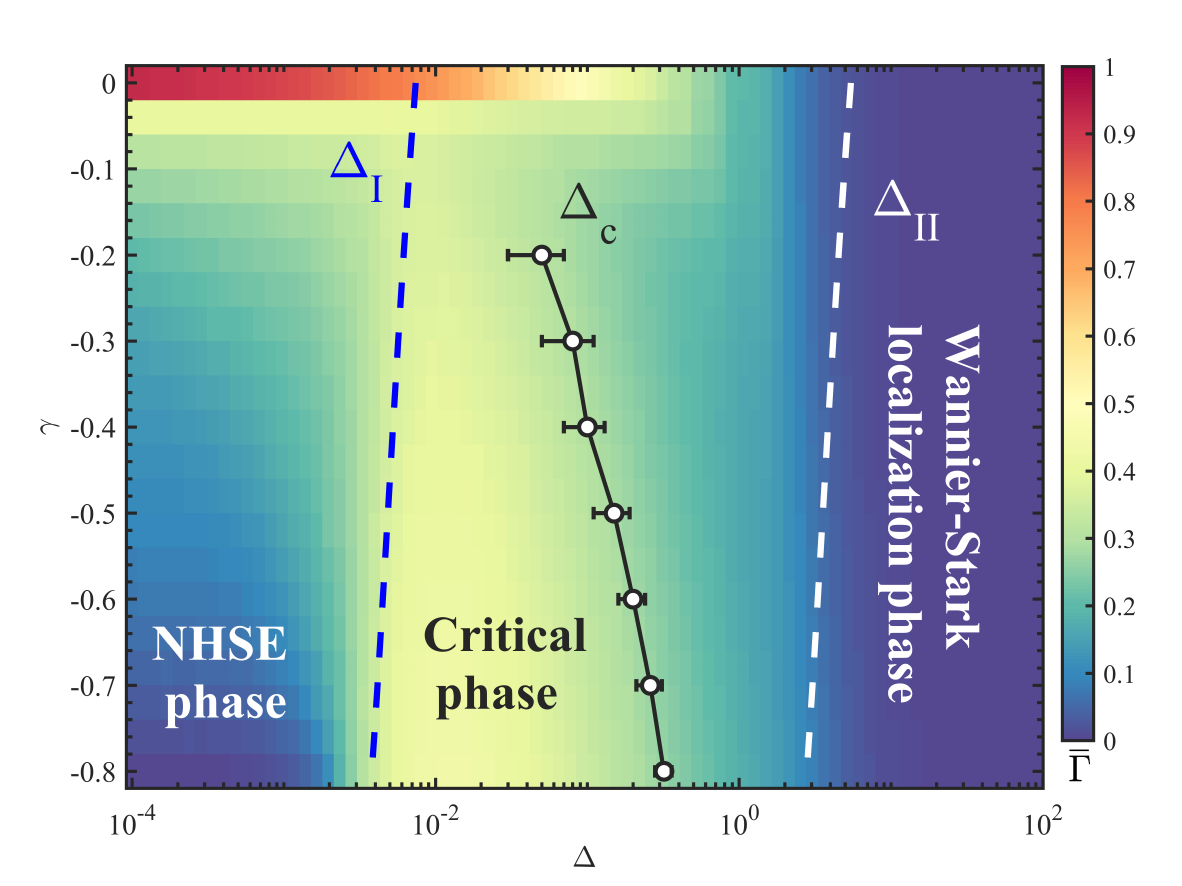}
	\caption{\label{fig:FD} The average fractal dimension for a system size $L=320$. The blue line, $\Delta_{\rm I} = 2e^{|g|+1}/L$, represents the boundary between the NHSE phase and the critical phase. The white dotted line, $\Delta_{\rm II} = 2e^{|g|}$, delineates the boundary between the critical phase and the Wannier-Stark localization phase. The black scatter plot, ${\Delta_c}$, indicates the algebra scaling line for non-Hermitian free fermions.}
\end{figure}

In this section, we will argue the entanglement behavior of non-Hermitian free-fermions with WSL from the perspective of single-particle localization. There has already been some discussion on the relationship between entanglement transitions and Anderson localization \cite{51,52,53}. To analytically argue the critical line $\{\Delta_c\}$ of Eq.~\ref{eq1} $H_{\rm eff}$ under OBC for the single-particle case, we first consider employing a similarity transformation $\mathcal{S}$ to obtain the Hermitian Hamiltonian $\mathcal{H}$ \cite{25},
\begin{align}
    \mathcal{H}=\mathcal{S}^{-1}H_{\rm eff}\mathcal{S}=
    \begin{pmatrix}
        F_1 & J' & & & & \\
        J' & F_2 & J' & & &\\
        & \ddots & \ddots & \ddots & \\
        & & \ddots & \ddots & J' &\\
        & & & J' & F_L&\\
    \end{pmatrix},
\end{align}
where $J'=\sqrt{J_RJ_L}$, $F_j=\Delta\cdot j$, the similarity matrix $\mathcal{S}={\rm diag}(e^{-g}, e^{-2g}, \cdots, -e^{Lg})$, and $g=\sqrt{J_R/J_L}$. Notably, the spectra of $ H_{\rm eff} $ and $ \mathcal{H} $ are identical under a similarity transformation. Substitute the state $\ket{\Psi'}=\sum_j\psi'_j\ket{j}$ into the eigenequation $\mathcal{H}\ket{\Psi'}=E\ket{\Psi'}$ and it satisfies $\ket{\Psi}=\mathcal{S}^{-1}\ket{\Psi'}$. Therefore, applying $\mathcal{S}^{-1}$ to an extended eigenstate of the Hamiltonian $H'$ results in the wave function becoming exponentially localized at the boundary. For a localized state, the wave function act as

\begin{align}
    |\psi_j| \propto\left\{ 
    \begin{array}{lc}
        e^{-(\tau(E)-g)(j-j_0)}, & j>j_0,\\
        e^{-(\tau(E)+g)(j_0-j)}, & j<j_0.\\
    \end{array}
    \right.
\end{align}
In this context, $j_0$ denotes the center of localization, and $\tau(E)>0$ represents the Lyapunov exponent associated with the Hamiltonian $H'$. Distinct Lyapunov exponents $\tau(E)\pm g$ are present on either side of the localization center. Delocalization occurs on one side when $\tau(E)\leq |g|$, and the critical point for transitioning from a localized state to the NHSE mode is $\tau(E)=|g|$. According to Ref.~\cite{68}, the analytical determination of whether the NHSE exists under OBC is governed by $\Delta_{\rm I}=2e^{|g|+1}/L$. To identify the mobility-edge boundaries, which demarcate the transition between delocalized and localized eigenstates, we can utilize wave function normalization. Given that a localized eigenstate remains unaffected by the system's boundary conditions, it follows that the threshold for the Wannier-Stark localization-delocalization transition under PBC is similarly defined by $\tau(E) = |g|$. Consequently, the non-Hermitian mobility-edge can be represented as $\Delta_{\rm II} = 2e^{|g|}$. $\Delta_{\rm I}$ [see FIG.~\ref{fig:FD} blue dotted line] and $\Delta_{\rm II}$ [see FIG.~\ref{fig:FD} white dotted line] indicate whether there is an occurrence or absence of NHSE and transitions between Wannier-Stark localization and delocalization.

Numerically, the shifts among the NHSE phase, the critical phase, and the Wannier-Stark localization phase can be characterized using the fractal dimension, denoted as $\Gamma$. For the $\alpha$-th eigenstate, expressed as $\ket{\Psi(\alpha)} = \sum_j \psi_j(\alpha) \ket{j}$, one can calculate the moments $\xi_q(\alpha) = \sum^L_{j=1} |\psi_j(\alpha)|^{2q} \propto L^{-\Gamma_q(q-1)}$ \cite{69,70,71}, where $\Gamma_q$ represents the fractal dimensions. By selecting $q=2$, the fractal dimension can be articulated as:
\begin{align}
\Gamma(\alpha) = \lim_{L \to \infty} \frac{\ln \xi(\alpha)}{\ln L}.
\end{align}
This expression provides a quantitative means to analyze transitions across different quantum phases numerically.
We omit the subscripts for $\Gamma_2$ and $\xi_2$, then $\xi$ is the inverse participation ratio. For an extended (localized) state $\Gamma=1$ $(\Gamma=0)$, and $0<\Gamma<1$ for a critical state. We introduce the average fractal dimension, denoted as $\bar{\Gamma} = \frac{1}{L} \sum^L_{\alpha=1} \Gamma(\alpha)$, to characterize the localization phase transition. As shown in FIG.~\ref{fig:FD}, when the value of $\Delta$ is less than $\Delta_{\rm I}$, the magnitude of $\bar{\Gamma}$ decreases with increasing $|\gamma|$, leading to a localized state, where NHSE plays a significant role. For $\Delta$ values between $\Delta_{\rm I}$ and $\Delta_{\rm II}$, the system enters a broad critical state region. When $\Delta > \Delta_{\rm II}$, the system immediately transitions to a Wannier-Stark localized state. Notably, our previous results of the critical scaling lines for dynamical entanglement transitions in non-Hermitian free fermions with a WSL, depicted by the black lines $\{\Delta_c\}$ in FIG.\ref{fig:FD}, reside within the critical phase region between NHSE phase and Wannier-Stark localization phase. Therefore, we suggest that the entanglement transitions for non-Hermitian free fermions with a WSL also correspond to the single-particle phase transition from NHSE states to Wannier-Stark localized states. This bears a strong resemblance to the argument for NHSE states to Anderson localization states under Anderson localization in non-Hermitian free fermions \cite{51,52,53}.

\section{Conclusion and outlooks}\label{conclusion_and_outlooks}

The results presented in this work highlight the nontrivial interplay between the WSL and NHSE. We have explored the non-unitary entanglement properties of one-dimensional non-Hermitian free fermions with WSL under both OBCs and PBCs.

Under OBCs, we convincingly demonstrate that non-Hermitian free fermions exhibit an algebraic phase between two area-law phases due to the introduction of WSL. By utilizing the steady state half-chain EE $S_{L/2}$ as a diagnostic tool, we reveal three distinct phases in the phase diagram: the area-law phase induced by the NHSE, the area-law phase induced by Wannier-Stark localization, and the intermediate algebraic scaling phase. The EE and mutual information further confirm the emergence of an algebraic scaling phase between the two area-law phases, a feature previously explored only in the context of Anderson localization \cite{51,52,53}. We applied the finite-size scaling method to determine the critical scaling properties of the algebraic scaling law. Under PBCs, the system shows only a transition from volume-law to area-law entanglement, lacking the NHSE and thus differing from the behavior seen in Anderson localization. We discussed the fundamental differences between the area-law phases induced by NHSE and Wannier-Stark localization from the perspective of single-particle localization. We also examined the potential impact of the critical scaling phase between the two distinct area-law phases on the entanglement behavior of non-Hermitian free fermions with WSL. We suggest that the emergence of this critical phase leads to the algebraic scaling law. Notably, the steady-state entanglement phase transitions induced by the combined effects of WSL and NHSE in this work also show interesting potential connections to the resonant enhanced tunneling (RET) effect, which we discuss in Appendix~\ref{RET}.

Our findings illuminate the dynamic entanglement behavior resulting from the interaction between the NHSE and WSL, uncovering new aspects of entanglement dynamics and phase transitions. Additionally, a promising avenue for future research would be to explore monitored fermions under the influence of WSL and to investigate the entanglement dynamics during quantum jumps, particularly when specific operations, such as feedback~\cite{RJ, XF1, XF2, ZCL}, are introduced.

\begin{acknowledgments}
We deeply appreciate Shan-Zhong Li, Shuo Liu and Ze-Chuan Liu for the valuable discussions and support, and H.-Z. Li thanks Wen Wang for her assistance on this work. H.-Z. Li also acknowledges Xue-Jia Yu for the guidance and support. Finally, we dedicate this paper to the vibrant and rapidly emerging Institute for Quantum Science and Technology (IQST) at Shanghai University. We gratefully acknowledges the National Natural Science Foundation of China (Grant No. 11874316), the National Basic Research Program of China (Grant No. 2015CB921103), and the Program for Changjiang Scholars and Innovative Research Team in University (Grant No. IRT13093).
\end{acknowledgments}



\appendix

\section{The effective non-Hermitian Hamiltonian and experimental realization}\label{nh-effective H}

In this appendix, we will discuss the possible path for realizing the effective Hamiltonian Eq.~\ref{eq1} from the perspective of the quantum trajectory approach \cite{78,79,80,81} and experimental realization. 

\subsection{Effective non-Hermitian Hamiltonian}
Firstly, we begin by considering a Markovian open quantum system, which is generally described by the Lindblad master equation:
\begin{align}
    \frac{d\rho}{dt} = -i[H, \rho] + \sum_k \left( L_k \rho L_k^\dagger - \frac{1}{2} \{L_k^\dagger L_k, \rho\} \right).
\end{align}
In this equation, $\rho$ denotes the density operator, $H$ represents the Hamiltonian governing the coherent dynamics, and $L_k$ are the jump operators that account for the interaction with the external environment. This master equation can be reformulated to highlight the effective non-Hermitian Hamiltonian: $\frac{d\rho}{dt} = -i(H_{\text{eff}}\rho - \rho H_{\text{eff}}^\dagger) + \sum_k L_k\rho L_k^\dagger.$ where the effective non-Hermitian Hamiltonian is defined as:
\begin{align}
    H_{\text{eff}} \equiv H - \frac{i}{2} \sum_k L_k^\dagger L_k.
\end{align}
The term $\sum_k L_k\rho L_k^\dagger$ describes each quantum trajectory subject to stochastic loss events. In this context, $L_k\sqrt{\delta t}$ can be interpreted as a measurement operator for a signal within the time interval $[t, t + \delta t]$, and $1 - iH_{\text{eff}}\delta t$ can be considered as a measurement operator for no signals. Under continuous monitoring and postselection of null measurement outcomes, quantum jumps become irrelevant, and the dissipative dynamics are described by the effective non-Hermitian Hamiltonian $H_{\text{eff}}$.

To derive the non-Hermitian free fermions with WSL as presented in Eq.~\ref{eq1}, we select the Hamiltonian $H$ and the jump operators $L_j$ (for $j = 1, 2, \ldots, L$) as follows:
\begin{align}
    H = -J \sum_{j=1}^{L-1} (c_{j+1}^\dagger c_j + c_j^\dagger c_{j+1}) + \sum_{j=1}^L F_j n_j,
\end{align}
where, $F_j \equiv \Delta \cdot j$.
\begin{align}
    L_j = \sqrt{|\gamma|} (c_j + i \, \text{sgn}(\gamma)c_{j+1}).
\end{align}
The effective Hamiltonian $H_{\text{eff}}$ derived from these choices differs from the original Hamiltonian by a background constant loss term $-i|\gamma| \sum_{j=1}^L n_j$, which describes the total decay of the system but does not contribute to the wave function dynamics. The resulting effective non-Hermitian Hamiltonian is:
\begin{align}
    H_{\text{eff}} = \sum_{j=1}^{L-1} (J_L c_j^\dagger c_{j+1} + J_R c_{j+1}^\dagger c_j) + \sum_{j=1}^{L} F_j n_j,
\end{align}
where the Hatano-Nelson NN hopping amplitudes are given by $J_L = -(1 - \gamma)$ and $J_R = -(1 + \gamma)$, with $\gamma$ representing the asymmetric non-Hermitian strength.

Open quantum dynamics, governed by non-Hermitian Hamiltonians, often face the challenge of a low success rate over time when trying to achieve the desired Hamiltonian. In contrast, the quantum master equation describes the dynamics of mixed states averaged over numerous quantum trajectories, thus avoiding the need for postselection. However, in specific scenarios, the issue of achieving an effective non-Hermitian Hamiltonian with a reasonable probability can be addressed \cite{72,73}. Similarly, in measurement-induced phase transitions, the experimental challenge is that only a quantum trajectory conditioned on specific measurement outcomes can exhibit an entanglement phase transition. The mixed quantum state averaged over multiple trajectories does not show such a phase transition. Recent proposals, however, suggest ways to realize measurement-induced phase transitions without the need for postselecting specific measurement outcomes \cite{74}. Finally, while this discussion centers on the quantum trajectory approach, the Feshbach projection formalism also provides a justification for effective non-Hermitian Hamiltonians \cite{75,76,77}.

\subsection{Experimental realization}

We propose an experimental setup for achieving Hatano-Nelson NN hopping and WSL. To implement Hatano-Nelson NN hopping, we suggest utilizing \textit{reservoir engineering}~\cite{reservoir1,reservior2}. Notably, previous studies, such as in Ref.~\cite{27}, have also explored cold-atom experiments in non-reciprocal hopping chains.

To experimentally realize the Hatano-Nelson NN hopping and WSL, we start by creating a one-dimensional optical lattice using counter-propagating laser beams. This setup forms a periodic potential where the lattice depth can be precisely controlled by adjusting the laser intensity~\cite{2007a,2007b,2009}. A linear potential gradient is then introduced by either applying a gravitational field or using a magnetic field gradient, inducing a linearly increasing potential across the lattice sites. This gradient results in a constant energy shift between neighboring sites, effectively creating the WSL~\cite{1996}. The atoms are initially cooled using laser cooling techniques, followed by evaporative cooling to reach ultracold temperatures~\cite{pra}. These ultracold atoms are then loaded into the optical lattice, occupying the lowest energy band. The linear potential gradient induces a Stark effect, shifting the energy levels and creating distinct energy states for each lattice site. The magnitude of the gradient and the depth of the lattice can be finely tuned to control the energy gaps, allowing for precise manipulation of the WSL structure.

To implement non-reciprocal hopping, we employ reservoir engineering. This involves introducing asymmetry in the dissipation channels~\cite{Metelmann2015}, achieved through light-induced transitions or coupling to a secondary atomic species. The asymmetry in dissipation can be controlled by varying the Rabi frequencies or the detuning of the coupling fields~\cite{Hafezi2011}, thus enabling directional hopping. It's crucial to maintain precise control~\cite{Fang2017} over these parameters to ensure the desired non-reciprocal effects.

The effective Hamiltonian can be decomposed into a Hermitian component, $\hat{H}_{\rm I}$, and an anti-Hermitian component, $\hat{H}_{\rm II}$. The Hermitian part is given by:
\begin{align}
\hat{H}_{\rm I} = -\sum_j[(a_{j+1}^\dagger a_j + \text{H.c.}) + F_j n_j],
\end{align}
which includes the WSL. The anti-Hermitian part, $\hat{H}_{\rm II}$, is described as:
\begin{align}
\hat{H}_{\rm II} = -\gamma\sum_j(a_{j+1}^\dagger a_j - \text{H.c.}),
\end{align}
where $\gamma$ characterizes the strength of the non-reciprocal hopping. This component can be realized through a jump operator $L_j = \sqrt{|\gamma|} (a_j + i \, \text{sgn}(\gamma)a_{j+1})$, which represents collective one-body loss. Here, $\text{sgn}(\gamma)$ indicates the sign of $\gamma$ and dictates the direction of the asymmetric hopping. In the absence of quantum jumps, the effective Hamiltonian governing the system's dynamics is:
\begin{align}
\hat{H}_{\rm eff} = \hat{H}_{\rm I} - \frac{i}{2}\sum_j L_j^\dagger L_j,
\end{align}
which can be further expanded to:
\begin{align}
\hat{H}_{\rm eff} = \hat{H}_{\rm I} + \hat{H}_{\rm II} - i\sum_j |\sinh{(\gamma)}|n_j,
\end{align}
where the term $-i\sum_j |\sinh{(\gamma)}|n_j$ corresponds to the loss of atoms on-site. Additionally, by controlling the boundaries at $j = 0$ and $j = L$, we can impose either PBC or OBC on the system. The experimental setup allows for detailed study of the dynamics of ultracold atoms under the influence of the WSL and the Hatano-Nelson effect. 

To further measure EE in this systems, we could employ \textit{quantum state tomography} (QST)~\cite{Banaszek1999,James2001,Haffner2005,Vogel1989,Paris2004}, which involves reconstructing the full density matrix of the system. The process can be broken down into the following:

\textit{Preparation and Initialization}: Prepare the quantum system in a well-defined initial state. For cold atoms in an optical lattice, this involves cooling the atoms to ultracold temperatures and loading them into the lattice. The system should be allowed to evolve under the influence of the Hamiltonian described previously, including the Hatano-Nelson NN hopping and WSL effects.

\textit{Measurement of Observables}: QST requires measuring a complete set of observables that correspond to different bases. For a system of $N$ qubits (or effective two-level systems), this involves performing measurements in various bases to obtain the probabilities of different outcomes. In practice, this can be done by applying a series of local unitary operations followed by a measurement in the computational basis. For cold atoms, this can be realized by manipulating the internal states of the atoms using laser pulses and then detecting the state populations using techniques like fluorescence imaging.

\textit{State Reconstruction}: The collected measurement data is used to reconstruct the density matrix $\rho$ of the quantum system. The density matrix provides a complete description of the quantum state, including all coherences and populations. The reconstruction can be performed using methods such as Maximum Likelihood Estimation (MLE), which finds the density matrix that best fits the observed data while ensuring physical properties like positivity and trace one are satisfied.

\textit{Computation of Reduced Density Matrix}: To find the entanglement entropy, we focus on a subsystem of interest. This requires tracing out the degrees of freedom associated with the rest of the system from the reconstructed density matrix $\rho$, yielding the reduced density matrix $\rho_A$ for the subsystem $A$. For example, if we are interested in the entanglement between two halves of the lattice, we would trace out the atoms in one half.

\textit{Calculation of Entanglement Entropy}: The entanglement entropy can be quantified using the von Neumann entropy, defined as:
   \begin{align}
   S(\rho_A) = -\text{Tr}(\rho_A \log \rho_A).
   \end{align}
   Alternatively, one can use the Rényi entropy as an approximation, which is given by:
   \begin{align}
   S_\alpha(\rho_A) = \frac{1}{1-\alpha} \log \text{Tr}(\rho_A^\alpha),
   \end{align}
   where $\alpha$ is a parameter. The case $\alpha = 1$ corresponds to the von Neumann entropy.

\textit{Error Analysis and Validation}: Due to experimental imperfections and statistical noise, it is crucial to assess the reliability of the reconstructed density matrix and the computed entanglement entropy. Techniques such as bootstrapping can be used to estimate the uncertainties in the measurements and to ensure that the results are statistically significant.

QST provides a comprehensive method for measuring entanglement entropy, but it is experimentally demanding due to the need for extensive measurements and precise control over the system. Despite these challenges, we still believe that it is possible to achieve this on the existing quantum simulation experimental platforms.


\section{Details on single-particle correlation matrix technique and entanglment entopy}\label{appendix_single-particle and EE}

In this section, we detail the methodology employed to compute the time-evolution and subsequent steady state EE in a non-Hermitian free fermions system from an arbitrary initial state utilizing a biorthogonal basis.

\subsection{single-particle correlation matrix technique}

We commence by considering an arbitrary initial state of the free fermion system represented by a Slater determinant, $\ket{\psi_0}=\prod^N_{i=1} c^{\dagger}_i \ket{\rm vac}$.
where $c^{\dagger}_i$ are the fermionic creation operators acting on the vacuum state $\ket{\rm vac}$, and $N$ denotes the number of particles in the system. The non-Hermitian Hamiltonian governing the system's dynamics is given by $H = \sum_{i,j}c_i^\dagger \mathcal{H}_{ij} c_j$. where $\mathcal{H}_{ij}$ are the elements of the non-Hermitian Hamiltonian matrix. Diagonalization of $H$ yields right and left eigenstates $\ket{\phi_i^R}$ and $\ket{\phi_i^L}$, corresponding to eigenvalues $\zeta_i$, such that:
\begin{align}
    H \ket{\phi_i^R} = \zeta_i \ket{\phi_i^R}\\
    H \ket{\phi_i^L} = \zeta^*_i \ket{\phi_i^L}
\end{align}
\begin{figure}[bt]
	\includegraphics[width=8.6cm]{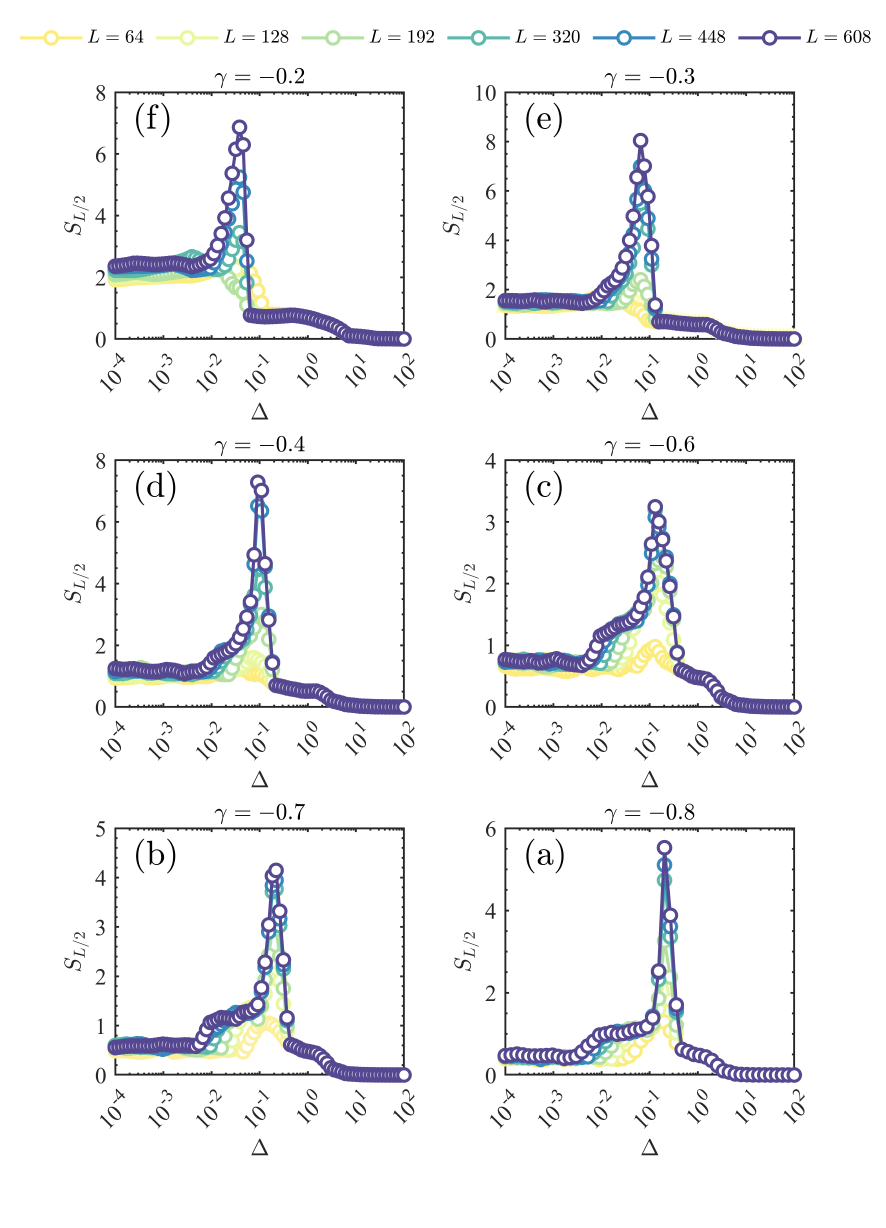}
	\caption{\label{fig:SL_datas} Steady-state half-chain entanglement entropy $S_{L/2}$ for the non-Hermitian effective Hamiltonian $H_{\rm eff}$, shown as a function of the WSL gradient $\Delta$. The data covers different non-Hermitian strengths $\gamma = -0.2$, $-0.3$, $-0.4$, $-0.6$, $-0.7$, $-0.8$, and various system sizes $L=64$, $128$, $320$, $448$, and $608$.}
\end{figure}

\begin{figure}[bt]
	\includegraphics[width=8.6cm]{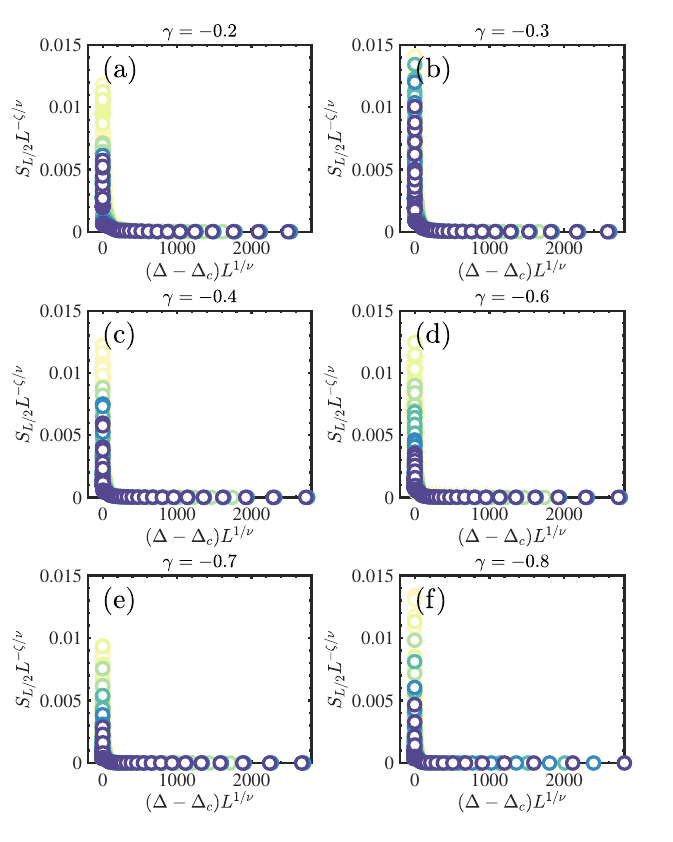}
	\caption{\label{fig:datacollapse} Data collapse of the steady-state half-chain entanglement entropy $S_{L/2}$ for the non-Hermitian effective Hamiltonian described by Eq.~\ref{eq1}, for various non-Hermitian strengths $\gamma = -0.2$, $-0.3$, $-0.4$, $-0.6$, $-0.7$, and $-0.8$. This collapse demonstrates the universal scaling behavior of $S_{L/2}$ across these different $\gamma$ values.}
\end{figure}

These satisfy the biorthogonal condition $\bra{\phi_i^L}\phi_j^R\rangle=\delta_{ij}$. Then, the temporal evolution of the state is conducted via the time-evolution operator $\ket{\psi(t)}=e^{-iHt}\ket{\psi_0}$. Expressing the time-evolution operator in the biorthogonal basis, we expand it as a sum over the right and left eigenstates:
\begin{align}
    e^{-iHt} = \sum_i e^{-i\zeta_i t}\ket{\phi_i^R}\bra{\phi_i^L}.
\end{align}
Applying Wick's theorem, we simplify the computation of the evolved state $\ket{\psi(t)}$ by decomposing it into contractions of single-particle operator pairs:
\begin{align}
    | \psi ( t ) \rangle = \prod _ { j = 1 } ^ { N } ( \sum _ { i = 1 } ^ { L } U _ { j i } ( t ) c _ { i } ) \ket{\rm vac} ,
\end{align}
where $U_{ij}(t)$ encapsulates the evolution of the single-particle states, and is determined by the matrix elements of $e^{-iHt}$ in the biorthogonal basis. The correlation matrix $C_{ij}(t)$ is then defined by the expectation values of the time-evolved operators:
\begin{align}
    C _ { i j } ( t ) = \langle \psi ( t ) | c _ { i } ^ { \dagger } ( t ) c _ { j } ( t ) | \psi ( t ) \rangle .
\end{align}

\subsection{Steady state EE in a non-Hermitian free fermions with Wannier-Stark ladder}

Given the time-evolution operator in the biorthogonal basis, $U(t) = e^{-iHt}$, which encapsulates the complex dynamics induced by the non-Hermitian Hamiltonian, we perform a QR decomposition at each time step $\Delta t$ to maintain numerical stability and accuracy. The QR decomposition is expressed as:
\begin{align}
    U ( t + \Delta t ) = Q ( t + \Delta t ) R ( t + \Delta t ) ,
\end{align}
where $Q(t+\Delta t)$ is an orthonormal matrix and $R(t+\Delta t)$ is an upper triangular matrix. This decomposition is crucial as it allows for a stable iterative update of the time-evolution matrix, particularly in the context of non-Hermitian systems where direct exponential operations can lead to numerical instabilities due to the unbounded growth or decay of matrix elements. The correlation matrix $C(t)$ at time $t$, which represents the pairwise correlations of fermionic operators within the system, is computed using the orthonormal matrix $Q(t)$:
\begin{align}
    C(t) = [Q(t)Q^\dagger(t)]^T.
\end{align}
This step is facilitated by the orthonormality of $Q(t)$, which ensures that $C(t)$ remains well-conditioned and suitable for further computations. The correlation matrix $C(t)$ reflects the occupancy probabilities of the single-particle states and is crucial for the evaluation of the subsystem's entanglement properties. To compute the EE $S_A$ for a subsystem $A$, we first extract the reduced correlation matrix $C_{ij}^A(t)$ corresponding to the subsystem by selecting the relevant rows and columns from $C_{ij}^A(t)$. We then calculate the eigenvalues $\{\lambda_k\}$ of $C_{ij}^A(t)$, each of which represents the occupation probability of mode $k$ in the subsystem. The von Neumann EE is given by:
\begin{align}
     S _ { A } = - \sum _ { k } \lambda _ { k } \log \lambda _ { k } + ( 1 - \lambda _ { k } ) \log ( 1 - \lambda _ { k } ) .
\end{align}
This entropy quantifies the amount of entanglement between subsystem $A$ and its environment.

In our numerical simulation, we set the time step size $\Delta t = 10$, the total number of time steps $N_t = 10000$. For Eq.~\ref{eq1}, the presence of a WSL introduces a weak Bloch oscillation, which result in significant noise in the EE. To address this, we consider smoothing the numerical results for the steady state EE by convolving them with a Gaussian $\mathcal{G}(n)=e^{-(n/\sigma)^2/2}$. 

\section{Steady state half-chain entanglment entropy for other values of $\gamma$}\label{appendix_engtanglement}
In this section, we present additional data to identify the critical scaling line ${\Delta_c}$ through the steady-state half-chain entanglement entropy (EE) for various values of $\gamma$ ($-0.2$, $-0.3$, $-0.4$, $-0.6$, $-0.7$, $-0.8$). As shown in the main text, FIG.~\ref{fig:SL_datas} illustrates the steady-state half-chain EE of the non-Hermitian effective model for $\gamma = -0.8$ (a), $\gamma = -0.7$ (b), $\gamma = -0.6$ (c), $\gamma = -0.4$ (d), $\gamma = -0.3$ (e), and $\gamma = -0.2$ (f) across different system sizes $L = 64$, $128$, $320$, $448$, $608$, plotted against the gradient of the Wannier-Stark ladders (WSL) $\Delta$. We observe that for all values of $\Delta$, $S_{L/2}$ displays distinct peaks as $\gamma$ varies, suggesting a size-dependent algebraic critical region.

\section{Data collapse for other values of $\gamma$}\label{appendix_datacollapse}

\begin{table}
\caption{\label{table1}Critical exponents and critical points of the effective Hamiltonian (Eq.~\ref{eq1}) for different values of $\gamma$.}
\begin{ruledtabular}
\begin{tabular}{cccc}
$\gamma$ & $\Delta_c$ & $\zeta$ & $\nu$ \\
\hline
-0.8 & 0.32 $\pm$ 0.04 & 2.12 $\pm$ 0.09 & 1.92 $\pm$ 0.06 \\
-0.7 & 0.26 $\pm$ 0.05 & 2.10 $\pm$ 0.09 & 1.85 $\pm$ 0.08 \\
-0.6 & 0.19 $\pm$ 0.04 & 1.96 $\pm$ 0.08 & 1.84 $\pm$ 0.08 \\
-0.5 & 0.15 $\pm$ 0.04 & 1.98 $\pm$ 0.06 & 1.92 $\pm$ 0.04 \\
-0.4 & 0.10 $\pm$ 0.03 & 2.04 $\pm$ 0.05 & 1.84 $\pm$ 0.06 \\
-0.3 & 0.08 $\pm$ 0.03 & 1.95 $\pm$ 0.06 & 1.87 $\pm$ 0.05 \\
-0.2 & 0.05 $\pm$ 0.02 & 2.09 $\pm$ 0.08 & 1.89 $\pm$ 0.06 \\
\end{tabular}
\end{ruledtabular}
\end{table}

In this section, we present additional data on the scaling collapse of the steady state half-chain EE $S_{L/2}$ for various values of $\gamma=~-0.2,~-0.3,~-0.4,~-0.6,~-0.7,~-0.8$ as shown in FIG.~\ref{fig:datacollapse}. The critical scaling exponents are also summarized in TABLE~\ref{table1}. We observe that the critical scaling exponent $\zeta$ and $\nu$ remain unchanged with increasing $\Delta$.

\section{Potential connections to the RET effect\label{RET}}

In this section, we will argue the interesting potential connection between the WSL and NHSE-induced steady-state entanglement phase transitions and the RET effect presented in the main text.

The RET effect is a critical phenomenon in quantum mechanics that examines how particles, such as electrons, can tunnel between different energy states within a system under resonance conditions. In experiments, Bose-Einstein Condensates (BEC) in optical lattices are often used to study this phenomenon. By manipulating external control parameters, such as the depth of the potential wells and acceleration-related parameters, resonant tunneling between different energy levels can be achieved, allowing for precise measurement of tunneling rates. Typically, these systems involve two primary energy levels, which can be adjusted according to experimental conditions. The system often exhibits the anticrossing phenomenon. Anticrossing occurs when two energy levels come very close to each other and strongly interact, causing their energies to avoid crossing and instead swap as the control parameters are varied. This phenomenon indicates that the energy levels do not simply intersect but rather repel each other, avoiding a direct crossing. Ref.~\cite{2007a} introduced that the anticrossing phenomenon is closely related to non-Hermitian Hamiltonians and the presence of gradient potential wells. The anticrossing not only affects the interaction between energy levels but also modifies the variation patterns of the tunneling rates between these levels, revealing more intricate dynamical behaviors.

We suggest that there might be an interesting connection between this phenomenon and our model Eq.~\ref{eq1}. It is possible that the acceleration and potential well depth in RET systems correspond to the nonreciprocal strength and WSL strength in our model. The anticrossing in RET systems may be related to the steady-state entanglement phase transitions discussed in this paper, which warrants further investigation.

\bibliography{bib.bib}

\begin{thebibliography}{116}
\bibitem [{are()}]{area_law}%
  \BibitemOpen
  \href@noop {} {}\bibinfo {note} {“Area law” states that EE is proportional to the boundary area of a subsystem, typically observed in 1d and 2d systems. "Volume law" indicates that EE scales with the volume of the subsystem, relevant when entropy growth exceeds simple boundary dependence. Even in 1d systems without a traditional "area," the area law still applies due to the presence of subsystem boundaries.}\BibitemShut {Stop}%
\bibitem [{den()}]{density_matrix}%
  \BibitemOpen
  \href@noop {} {}\bibinfo {note} {The biorthogonal density matrix is defined by $\rho=\sum_i |\phi_i^R\rangle \langle \phi_i^L|$}\BibitemShut {NoStop}%
\bibitem [{sca()}]{scaling}%
  \BibitemOpen
  \href@noop {} {}\bibinfo {note} {We use finite-size scaling and fitting methods to compare the critical point's scaling exponent, aiming to obtain more accurate critical information.}\BibitemShut {Stop}%
\bibitem [{ave()}]{average}%
  \BibitemOpen
  \href@noop {} {}\bibinfo {note} {It should be noted that the average critical exponents reflect the inherent properties of the system but do not correspond to any known universal class.}\BibitemShut {Stop}%
\end{thebibliography}

\end{document}